\documentclass[12pt]{article}

\usepackage{newtxtext,newtxmath}

\usepackage{graphicx}
\usepackage[version=4]{mhchem}
\usepackage{amsmath}
\usepackage{bm}
\usepackage[letterpaper,margin=1in]{geometry}

\renewenvironment{abstract}
	{\quotation}
	{\endquotation}

\date{}

\makeatletter
\renewcommand{\fnum@figure}{\textbf{Figure \thefigure}}
\renewcommand{\fnum@table}{\textbf{Table \thetable}}
\makeatother

\usepackage{scicite}

\usepackage{url}

\def\scititle{
	Orientation-resolved tunnel spectroscopy\\ of a triplet superconductor
}
\title{\bfseries \boldmath \scititle}

\author{
	Marcel Strohmeier$^{1\ast}$,
	Andriy Smolyanyuk$^{2}$,
    Karsten Held$^{2}$,
	Michael Smidman$^{3\dagger}$,\\
    Geetha Balakrishnan$^{3}$,
    Wolfgang Belzig$^{1}$,
    Elke Scheer$^{1\ast}$,
	Angelo Di Bernardo$^{1\ast\ddagger}$\and
	\small$^{1}$Department of Physics, University of Konstanz, 78457 Konstanz, Germany.\and\
	\small$^{2}$Institute of Solid State Physics, TU Wien, 1040 Vienna, Austria.\and
    \small$^{3}$Department of Physics, University of Warwick, Coventry CV4 7AL, United Kingdom.\and
	\small$^\ast$Corresponding author. Email: marcel.strohmeier@uni-konstanz.de,\\\small elke.scheer@uni-konstanz.de, angelo.dibernardo@uni-konstanz.de\and
	\small$^\dagger$Present address: Center for Correlated Matter and School of Physics, \\\small Zhejiang University, Hangzhou 310058, China.\and
    \small$^\ddagger$Present address: Department of Physics, University of Salerno, Fisciano SA 84084, Italy.
}

\begin{document} 

\maketitle

\begin{abstract} \bfseries \boldmath
\noindent 

Although Cooper pairs in superconductors generally condense into a spin-singlet state, spin-triplet superconductivity has attracted sustained interest since it enables dissipationless spin transport and is an important requirement for manifold topological quantum technologies. 
Noncentrosymmetric superconductors provide a promising route to triplet pairing because their intrinsic antisymmetric spin-orbit coupling (ASOC) can mix spin-singlet and spin-triplet states. To date, the strongest evidence for such mixed-parity superconductivity has been reported in heavy-fermion systems, where strong electronic correlations obscure the role of ASOC. Whether ASOC alone can support a substantial spin-triplet component remains unresolved. Here we show that the weakly correlated noncentrosymmetric superconductor \ce{Nb18Re82} (Nb-Re) hosts a mixed-parity superconducting state with a substantial spin-triplet contribution. Orientation-resolved low-temperature scanning tunneling spectroscopy reveals a strongly anisotropic quasiparticle density of states. A symmetry-constrained analysis shows that the spectroscopic dataset is explained by an order parameter combining a nodal spin-singlet component with a spin-triplet contribution reaching up to half of the singlet amplitude. These findings provide a momentum-resolved determination of a mixed-parity superconducting order parameter and demonstrate that ASOC alone can foster a substantial spin-triplet component. 
More broadly, they establish orientation-resolved tunneling spectroscopy as a route to identifying mixed-parity superconducting states and suggest that triplet superconductivity may be more widespread among noncentrosymmetric materials than previously recognized.

\end{abstract}

\section*{Introduction}

\noindent Spin-triplet superconductivity enables dissipationless spin transport through ferromagnets and represents a promising route towards fault-tolerant quantum computing \cite{Bergeret2005, Bernardo2015, Linder2015, Diesch2018, Sato2017, Kezilebieke2020, Qiu2021}. Among the materials supporting intrinsic spin-triplet Cooper pairs, noncentrosymmetric superconductors (NCSs) are particularly attractive because antisymmetric spin-orbit coupling (ASOC), arising from their broken inversion symmetry, naturally mixes spin-singlet and spin-triplet pairing states in these materials \cite{Yip2014, Kneidinger2015, Smidman2017, Sigrist2007}. 

The strongest experimental evidence for parity-mixed superconductivity in NCSs so far has been obtained in heavy-fermion materials \cite{Bauer2004, Yogi2004, Yogi2006}, where, nonetheless, strong electronic correlations obscure the role of ASOC towards the stabilization of spin-triplet superconductivity. Whether ASOC \emph{per se} can generate a substantial spin-triplet component in a weakly correlated NCS represents an outstanding question, for which existing evidence from bulk, orientation averaged probes has remained inconclusive ~\cite{Bao2013,Kim2018,Xu2014,Datta2021}. In Li$_2$Pt$_3$B, a singlet-to-triplet amplitude ratio
comparable to that identified here for Nb-Re has been deduced from several averaged probes on bulk samples without accessing the $k$-space structure of the order parameter ~\cite{Bao2013, Yuan2006}. In the topological semimetal YPtBi, a mixed singlet-septet pairing state has been proposed from measurements of the London penetration depth ~\cite{Kim2018}. This scenario arises from the $j=3/2$  character of a strongly spin-orbit-split semimetallic band structure, a qualitatively different pairing mechanism than the spin-$1/2$ singlet-triplet mixing in the present work. A claim of unconventional superconductivity has been put forward for the half-Heusler semimetal LuPdBi without addressing the singlet-triplet ratio ~\cite{Xu2014}. A local spectroscopic claim of mixed singlet-triplet pairing based on STS has been reported in the weakly correlated NCS Ru$_7$B$_3$ ~\cite{Datta2021}. In this study, however, tunneling spectra were acquired at randomly chosen points on crystallites of random crystallographic orientation, which cannot access the $k$-space structure of the order parameter. To our knowledge, a direct $k$-resolved reconstruction of a mixed-parity order parameter has not previously been achieved yet in any weakly correlated superconductor. 

Weakly correlated Re-based superconductors with the noncentrosymmetric $\alpha$-Mn structure are ideally suited to address these questions \cite{Kushwaha2024, Biswas2012, Singh2017, Shang2018, Singh2018}. Among them, \ce{Nb18Re82} (Nb-Re) is particularly attractive because it combines a high superconducting transition temperature $T_{c}$ ($>8.8$~K) compared to other NCSs \cite{Karki2011,Savitskii1966}, with compatibility for thin-film growth and superconducting device fabrication \cite{Koch2024, Cirillo2024, DeChiara2026}.
At the same time, Nb-Re has attracted considerable interest because previous studies have reported apparently conflicting experimental results consistent with either conventional Bardeen-Cooper-Schrieffer (BCS) theory or unconventional superconductivity. The latter type of evidence includes anomalously large upper critical fields $B_{c2}$ \cite{Kakhaki2023}, double-gap behavior \cite{Cirillo2015}, time-reversal symmetry breaking \cite{Shang2018}, and transport properties consistent with spin-triplet superconductivity in hybrids with ferromagnets \cite{Cirillo2025}. In contrast, bulk transport and thermodynamic measurements have suggested that Nb-Re is a weakly correlated, phonon-mediated NCS with a BCS-like superconducting OP \cite{Karki2011, Chen2013}. Overall, these results suggest that a direct determination of the symmetry of the superconducting OP in a NCS like Nb-Re requires a local, orientation sensitive probe capable of resolving its $k$-space structure, rather than non-local or orientation averaged measurements.

Here we combine low-temperature scanning tunneling microscopy/spectroscopy (low-$T$ STM/STS) with crystallographic orientation control, to reconstruct the superconducting OP of Nb-Re in $k$-space. 
By comparing tunneling spectra measured on Nb-Re thin films and single crystals exposing four distinct crystallographic surfaces, we identify a mixed-parity superconducting state comprising both a nodal spin-singlet and a substantial spin-triplet component. 

Our results demonstrate that ASOC alone is sufficient to generate substantial parity mixing in a weakly correlated NCS, thus establishing that intrinsic spin-triplet superconductivity is not only restricted to strongly correlated NCSs. More broadly, our experimental approach establishes a framework for the rational discovery and design of other quantum materials with intrinsic spin-triplet pairing.

\section*{Results}\label{sec2}

\subsection*{Multiple order parameter signatures in Nb-Re thin films}\label{subsec2}

\begin{figure}[t]
    \centering
    \includegraphics[width=1.0\columnwidth]{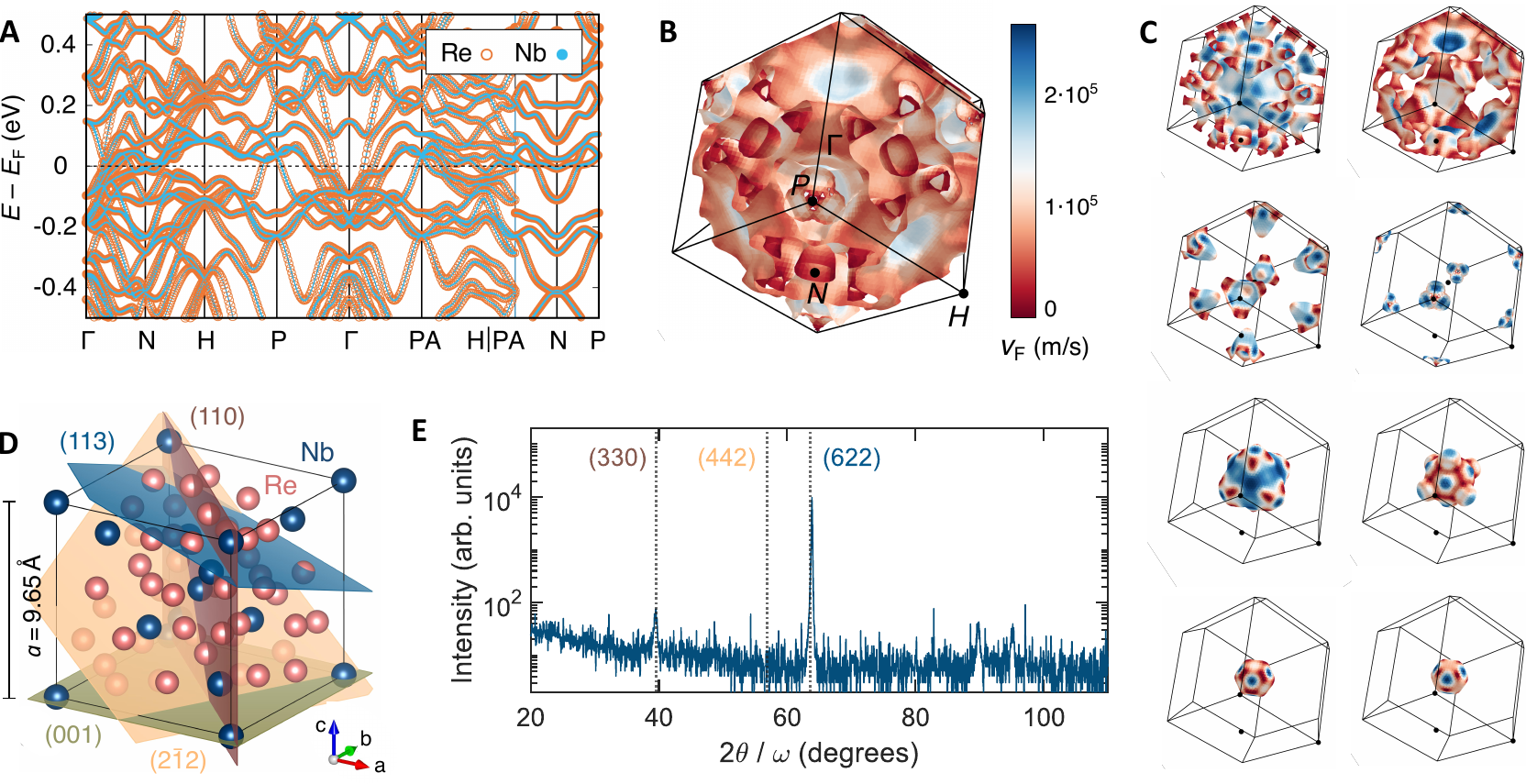}
    \caption{\textbf{Electronic structure and crystallographic characterization of Nb-Re.} (\textbf{A}) Electronic band structure based on density functional theory calculations. (\textbf{B}) Corresponding three-dimensional Fermi surface (FS) constructed from the band structure, illustrating the complex multi-band topology. The color scale reflects the magnitude of the Fermi velocity $v_{\text{F}}$. (\textbf{C}) FS decomposed into eight individual band contributions. (\textbf{D}) Cubic $\alpha$-Mn unit cell of Nb-Re with the colored crystallographic planes indicating the investigated surface orientations. (\textbf{E}) Single-crystal X-ray diffraction pattern measured on a (113)-oriented single crystal, confirming the crystallographic quality and orientation of the sample.}
    \label{Fig1}
\end{figure}

\noindent Nb-Re crystallizes in the noncentrosymmetric cubic $\alpha$-Mn structure (space group $I\overline{4}3m$), with a large unit cell (lattice constant of 9.651\,\AA) that results in pronounced differences in atomic packing between different crystallographic planes (Fig.~\ref{Fig1}D). Density functional theory (DFT) calculations also reveal a complex multiband electronic structure with two bands dominating the density of states (DoS) near the Fermi level (Figs.~\ref{Fig1}A-C). The pronounced structural anisotropy of Nb-Re makes it an ideal system for orientation resolved tunneling spectroscopy, where different crystallographic surfaces probe distinct cross-sections of the superconducting OP in $k$-space. If the OP has an anisotropic $k$-space structure, crystal surfaces with different orientations should in fact produce distinct tunneling spectra.

\begin{figure}[t]
    \centering
    \includegraphics[width=1.0\columnwidth]{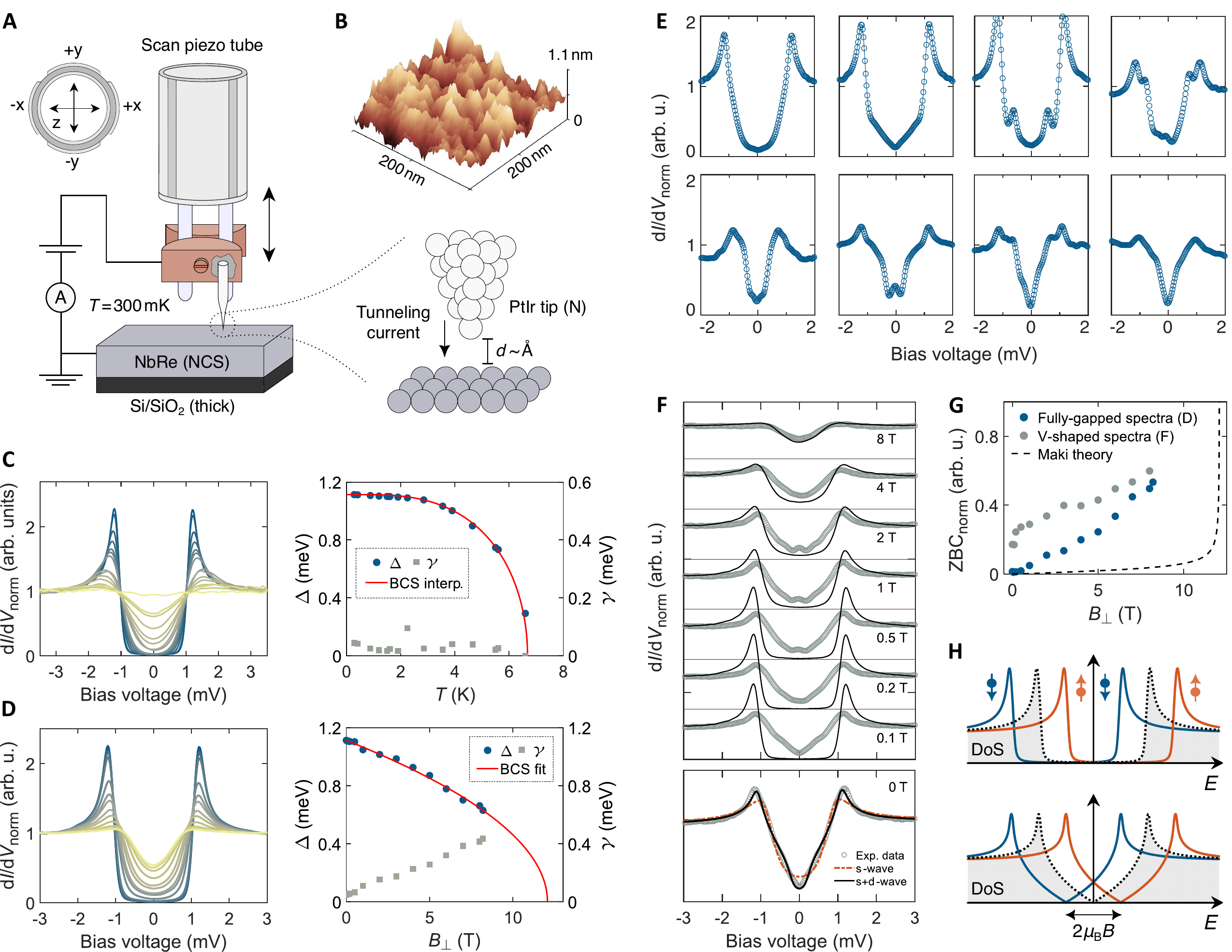}
    \caption{\textbf{Tunneling spectroscopy on Nb-Re thin films.} (\textbf{A}) Scheme of the low-temperature STM experiment and the normal metal/superconductor tunneling configuration. (\textbf{B}) Three-dimensional representation of an atomic force microscope image showing the surface topography of a Nb-Re thin film. (\textbf{C}), (\textbf{D}) BCS-like tunneling spectra measured at various temperatures $T$ and perpendicular magnetic fields $B_\perp$. The extracted OP 
    and Dynes parameter $\gamma$ are shown on the right. (\textbf{E}) Selection of non-BCS-type d$I$/d$V$ spectra measured across the Nb-Re surface. (\textbf{F}) V-shaped tunneling spectrum with its evolution under an increasing $B_\perp$. (\textbf{G}) Zero-bias conductance (ZBC) as a function of $B_\perp$. (\textbf{H}) Illustration of the spin-split DoS with $s$-wave (upper panel) and $d$-wave (lower panel) OP in the presence of a magnetic field.}
    \label{Fig2}
\end{figure}

To determine whether the superconducting OP is anisotropic, we first performed low-$T$ STM/STS on polycrystalline Nb-Re thin films. Because the grain size of these films ($\approx5-10$\, nm) is comparable to both the superconducting coherence length ($\xi \approx 4-5$\ \,nm for disordered Nb-Re thin films \cite{Karki2011, Cirillo2016}) and to the lateral resolution of our STM (Fig.~\ref{Fig2}B) \cite{Debuschewitz2007}, individual tunneling spectra predominantly probe single crystallites with well-defined, albeit unknown, crystallographic orientation. These measurements provide a statistical ensemble of local spectra acquired from differently oriented grains. 

The differential conductance (d$I$/d$V$) spectra $-$ proportional to the superconducting DoS $-$ measured by low-$T$ STS on Nb-Re thin films reveal two distinct types of behavior. In some regions, the spectra are fully gapped and can be quantitatively described by the Dynes model, 
with a superconducting OP whose temperature ($T$) and magnetic-field ($B$) dependence follow those expected for a conventional BCS superconductor (Figs.~\ref{Fig2}C,\,D). Elsewhere, the spectra exhibit pronounced deviations from isotropic behavior, including V-shaped low-energy DoS with split 
peaks at the gap edge, or a slope change of the DoS in the gap region with wider 
peaks (Figs.~\ref{Fig2}E). These spectral features are indicative of a nodal structure of the superconducting OP.

Further evidence for nodal superconductivity comes from the $B$ dependence of the zero-bias conductance (ZBC). For V-shaped spectra, 
the ZBC increases immediately upon $B$ application and subsequently grows approximately linearly with $B$ (Figs.~\ref{Fig2}F,\,G), consistent with the presence of low-energy quasiparticle states associated with nodes in the superconducting OP. By contrast, spectra exhibiting a 
BCS-like DoS show no such immediate response at low $B$. This contrasting behavior, illustrated schematically in Fig.~\ref{Fig2}H, reflects the presence of quasiparticles already at $B$ = 0, which are immediately redistributed by an applied $B$, in contrast with a fully gapped superconductor where a finite $B$ is required to generate low-energy quasiparticle states. 

Taken together, these results demonstrate that the 
quasiparticle DoS depends strongly on crystallographic orientation, providing direct evidence for a highly anisotropic superconducting gap structure. This conclusion naturally motivates orientation resolved spectroscopic measurements on Nb-Re single crystals with well defined crystallographic surfaces, establishing a direct link 
between the tunneling spectra and the $k$-dependence of the superconducting OP.

\subsection*{Orientation dependent DoS in Nb-Re single crystals}\label{subsec3}
\begin{figure}[t]
    \centering
    \includegraphics[width=1.0\columnwidth]{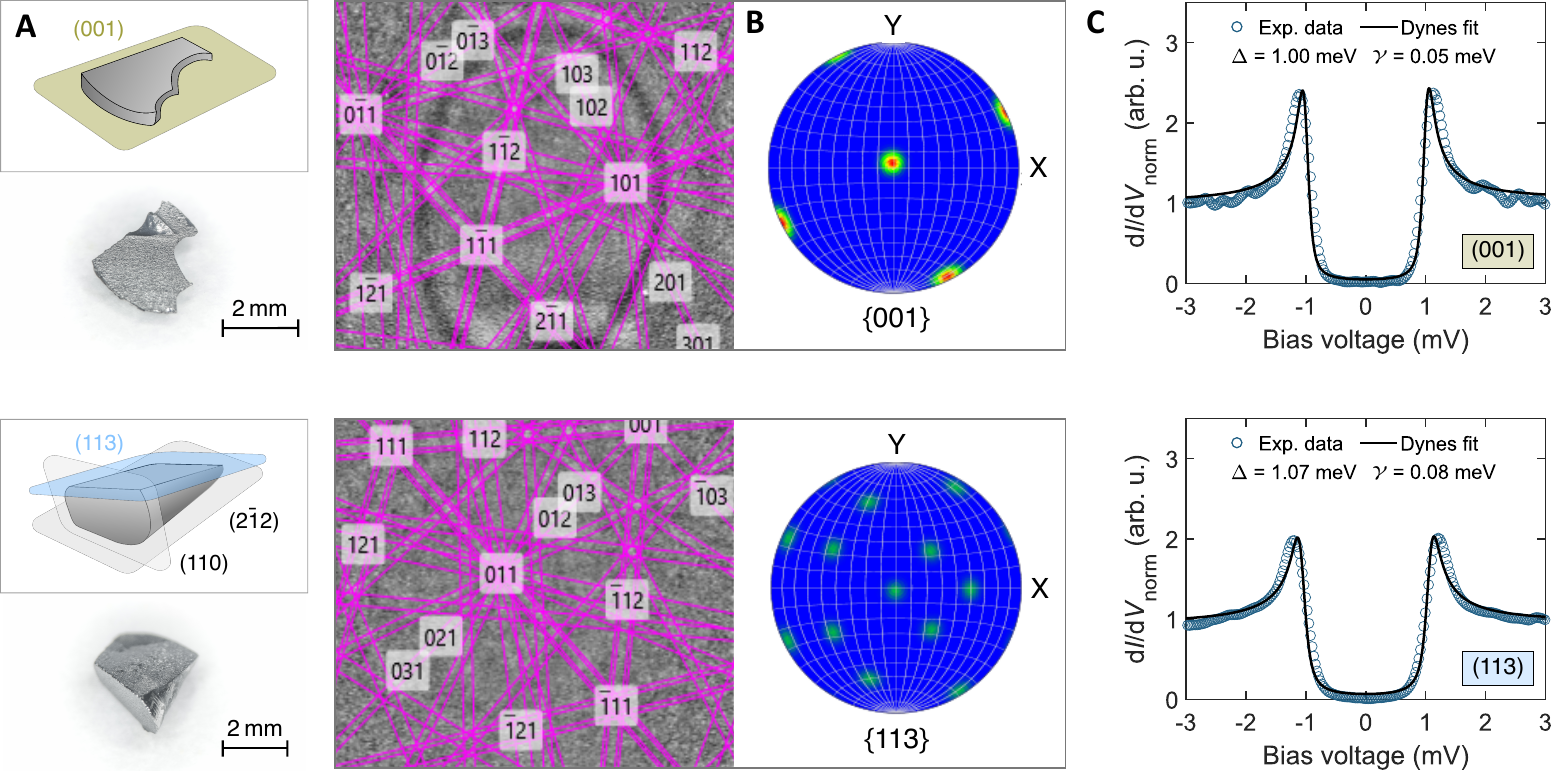}
    \caption{\textbf{Characterization of Nb-Re single crystals with (001) and (113) surface orientations.} (\textbf{A}) Photographs of the two Nb-Re single crystals, along with sketches of the sample geometry illustrating the investigated crystallographic surface orientations. (\textbf{B}) Electron backscatter diffraction characterization showing the Kikuchi patterns (left) and the corresponding pole figures (right) for the two colored surface orientations in (A). (\textbf{C}) Representative BCS-type tunneling spectra acquired on the (001) and (113) surfaces.}
    \label{Fig3}
\end{figure}

\noindent 
Although the thin-film measurements reveal orientation dependent spectroscopic signatures, the unknown crystallographic orientation of individual grains prevents a direct assignment of tunneling spectra to specific crystallographic orientations. To resolve this ambiguity,
we performed low-$T$ STM/STS on Nb-Re single crystals exposing four crystallographic surfaces, whose orientations were confirmed by electron back-scattered diffraction and X-ray diffraction (Figs.~\ref{Fig1}E and \ref{Fig3}B, and Supplementary Fig. S1).

The (001)- and (113)-oriented Nb-Re single crystals (Fig.~\ref{Fig3}A) exhibit 
BCS-like DoS with a gap amplitude $\Delta \approx 1.0$\,meV  at $T = 300$ mK (Fig.~\ref{Fig3}C), consistent with the 
spectra measured on some regions of the Nb-Re thin films (Figs.~\ref{Fig2}C,\,D). In contrast, the (2$\overline{1}$2)-oriented single crystal displays 
BCS-like as well as the full range of non-BCS spectral signatures identified in the thin films $-$ including double-peak features, V-shaped gaps, and split peaks at the gap edge (Fig.~\ref{Fig4}A). The observation of such spectra also on a single crystal 
demonstrates that these spectral features are intrinsic and cannot be attributed to averaging over differently oriented grains in the Nb-Re thin films. Furthermore, the double-peak spectra remain perfectly symmetric with respect to bias polarity under applied $B$, which rules out impurity induced bound states like Yu-Shiba-Rusinov \cite{Yazdani1997} states 
(Fig.~\ref{Fig4}B).

Additional evidence for an anisotropic superconducting state is provided by the $B$-response of the spectra measured on the (2$\overline{1}$2) surface. At a given $B$, a magnetic field applied parallel to the surface ($B_\parallel$) produces more V-shaped spectra than a field applied perpendicular to the surface ($B_\perp$) (Fig.~\ref{Fig4}C). This anisotropic behavior is consistently reproduced across different sample regions and provides independent evidence for a superconducting state with a highly anisotropic $k$-space structure. 

The (110)-oriented single crystal exhibits a similar variety of spectral features with a stronger tendency towards V-shaped tunnel spectra. In addition, on this surface we find an enlarged $\Delta$ $\sim 2$\,meV alongside ZBC peaks (Fig.~\ref{Fig4}D) consistent with surface induced Andreev bound states \cite{Hu1994, Kashiwaya1995}. 
Further details and a quantitative cluster based analysis of the spectral variation measured on the (110) and (2$\overline{1}$2) Nb-Re single-crystal surfaces are provided in the Supplementary Materials.

\subsection*{Symmetry-constrained order parameter model}\label{subsec4}

\noindent The orientation dependent spectra shown above impose stringent constraints on the symmetry of the superconducting OP. Any viable model explaining the OP symmetry must simultaneously reproduce (i) 
BCS-like spectra on the (001) and (113) surfaces, (ii) V-shaped low-energy DoS indicative of nodes, and (iii) double-peak spectral structures observed on the (110) and (2$\overline{1}$2) surfaces. We therefore examine the superconducting states permitted by the crystal symmetry of Nb-Re and determine those consistent with the complete experimental dataset.

\begin{figure}[t]
    \centering
    \includegraphics[width=1.0\columnwidth]{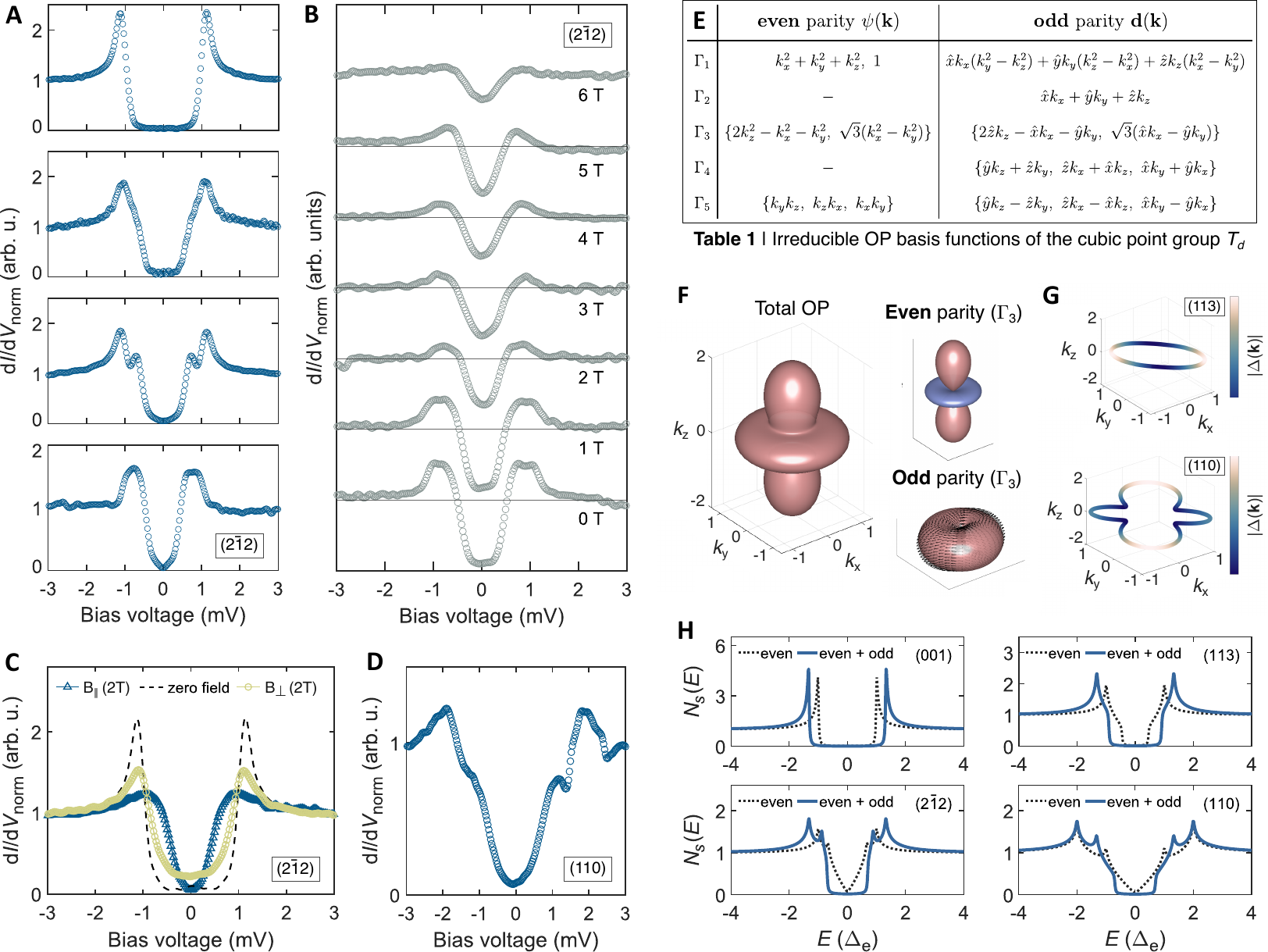}
    \caption{\textbf{Tunneling spectroscopy on a Nb-Re single crystal with (2$\overline{1}$2) surface orientation.} (\textbf{A}) Representative d$I$/d$V$ tunnel spectra measured on the ($2\overline{1}2$) surface of a Nb-Re single crystal. (\textbf{B}) Field-dependent evolution of a double-peak spectrum under a magnetic field $B_\perp$. (\textbf{C}) Anisotropic suppression of the OP $-$ determined from the spacing of the coherence peaks $-$ on the ($2\overline{1}2$) surface, with d$I$/d$V$ measured consecutively in a 2\,T magnetic field applied perpendicular and parallel to the sample surface, respectively. (\textbf{D}) Enlarged OP observed on the (110) surface orientation. (\textbf{E}) Summary of the even- and odd-parity irreducible OP basis functions compatible with the  point group of Nb-Re. (\textbf{F}) Three-dimensional illustration of the total superconducting OP, constructed from the irreducible basis functions compatible with \ce{Nb18Re82}. A combination of an even-parity spin-singlet and an odd-parity spin-triplet component is required to account for the full set of d$I$/d$V$ spectra acquired on four different crystallographic surface orientations. (\textbf{G}) Corresponding two-dimensional (2D) OP profiles along two exemplary planes, (110) and (113). (\textbf{H}) Model-based calculations of the DoS corresponding to specific direction-dependent 2D OP shown in (F).}
    \label{Fig4}
\end{figure}

The superconducting OP in a NCS with ASOC mixing spin-singlet and spin-triplet components can be generally written as 
\begin{equation}
\hat{\Delta}(\mathbf{k}) = \left[ \Delta_e \, \psi(\mathbf{k}) + \Delta_o \, \mathbf{d}(\mathbf{k}) \cdot \hat{\bm{\sigma}} \right] i \sigma_2 \;, 
     \label{eq:OrderParameter}
 \end{equation}
where $\hat{\bm{\sigma}}=\sigma_1\,\hat{\textbf{\textit{x}}}_1 + \sigma_2\,\hat{\textbf{\textit{x}}}_2 + \sigma_3\,\hat{\textbf{\textit{x}}}_3$ is the Pauli vector, $\psi(\mathbf{k})$ and $\mathbf{d}(\mathbf{k})$ are the even-parity spin-singlet and odd-parity spin-triplet basis functions, and $\Delta_e$ and $\Delta_o$ are their respective amplitudes. To identify the superconducting state consistent with the experimental data, we evaluate all the basis functions allowed by the $T_d$ point group of Nb-Re (Fig.~\ref{Fig4}E), and calculate the corresponding superconducting DoS for each crystallographic orientation. As a minimal description, we assume a spherical Fermi surface (FS), and compute the superconducting DoS by integrating over the momentum states parallel to the exposed crystal plane,  $\mathbf{k}_\parallel$ $-$ which correspond to a two-dimensional cross-section of the FS along the same plane (see Supplementary Text for a detailed discussion of the tunneling model). The superconducting DoS as a function of energy ($N_s(E)$) is given by 

 \begin{equation}
     N_s(E) = \left\langle \text{Re} \left[ \frac{|E|}{\sqrt{E^2 - |{\Delta}(\mathbf{k})|^2}} \right] \right\rangle_{\mathbf{k}_\parallel,\,\mathrm{FS}}.\vspace{0.3cm}
     \label{eq:DoS}
 \end{equation}


To reproduce the measured d$I$/d$V$ spectra, we first consider the simplest possible superconducting state: an isotropic $s$-wave OP belonging to the fully symmetric ($\Gamma_1$) representation. Although this state reproduces the fully gapped spectra observed on the (001) and (113) surfaces, it cannot account for the V-shaped DoS, double-peak structures, or ZBC features observed on the (110) and (2$\overline{1}$2) surfaces, which suggests that the superconducting OP must possess a non-trivial $k$-dependence.

Among all even-parity 
basis functions allowed by $T_d$ symmetry (Fig.~\ref{Fig4}E; left column), only the $d_z^2$ function of $\Gamma_3$ produces the combination of nodal spectra and unequal gap amplitudes required to reproduce the V-shaped spectra and double-peak structures observed experimentally. As shown in Fig.~\ref{Fig4}H (dashed curves), this state naturally yields nodal spectra for selected surface orientations and fully gapped spectra for other orientations, in agreement with experiment.

Despite this agreement, the $\Gamma_3$ spin-singlet state alone cannot reproduce the complete spectroscopic dataset. In particular, it fails to account for the double-peak spectra without a V-shaped background observed on the (2$\overline{1}$2) and (110) surfaces. Reproducing these spectra requires the inclusion of an additional 
spin-triplet component (Fig.~\ref{Fig4}E; right column).

To obtain a mixed-parity ${\Delta}(\mathbf{k})$, 
we define $\Delta_\text{even}=\Delta_e|\psi(\mathbf{k})|$ and $\Delta_\text{odd}=\Delta_o|\mathbf{d}(\mathbf{k})|$, and combine them as $|{\Delta}(\mathbf{k})|$ $=$ $\sqrt{(\Delta_\text{even})^2 + (\Delta_\text{odd})^2}$. This minimal description neglects any interference terms between the singlet and triplet channels and the helicity-band splitting generated by ASOC (see Supplementary Text for details). We find that the complete experimental dataset can be reproduced only by including an odd-parity spin-triplet component with toroidal shape belonging to the same ($\Gamma_3$) representation as the even-parity spin-singlet $d_z^2$ (Fig.~\ref{Fig4}F). As shown in Fig.~\ref{Fig4}H (solid blue curves), this mixed-parity superconducting OP reproduces the full range of experimental spectra including fully gapped 
spectra along \{001\} or \{113\}, 
 double-peak spectra along \{110\} and \{2$\overline{1}$2\}, and nodal V-shaped spectra when the triplet component vanishes.

\begin{figure}[t]
    \centering
    \includegraphics[width=1.0\columnwidth]{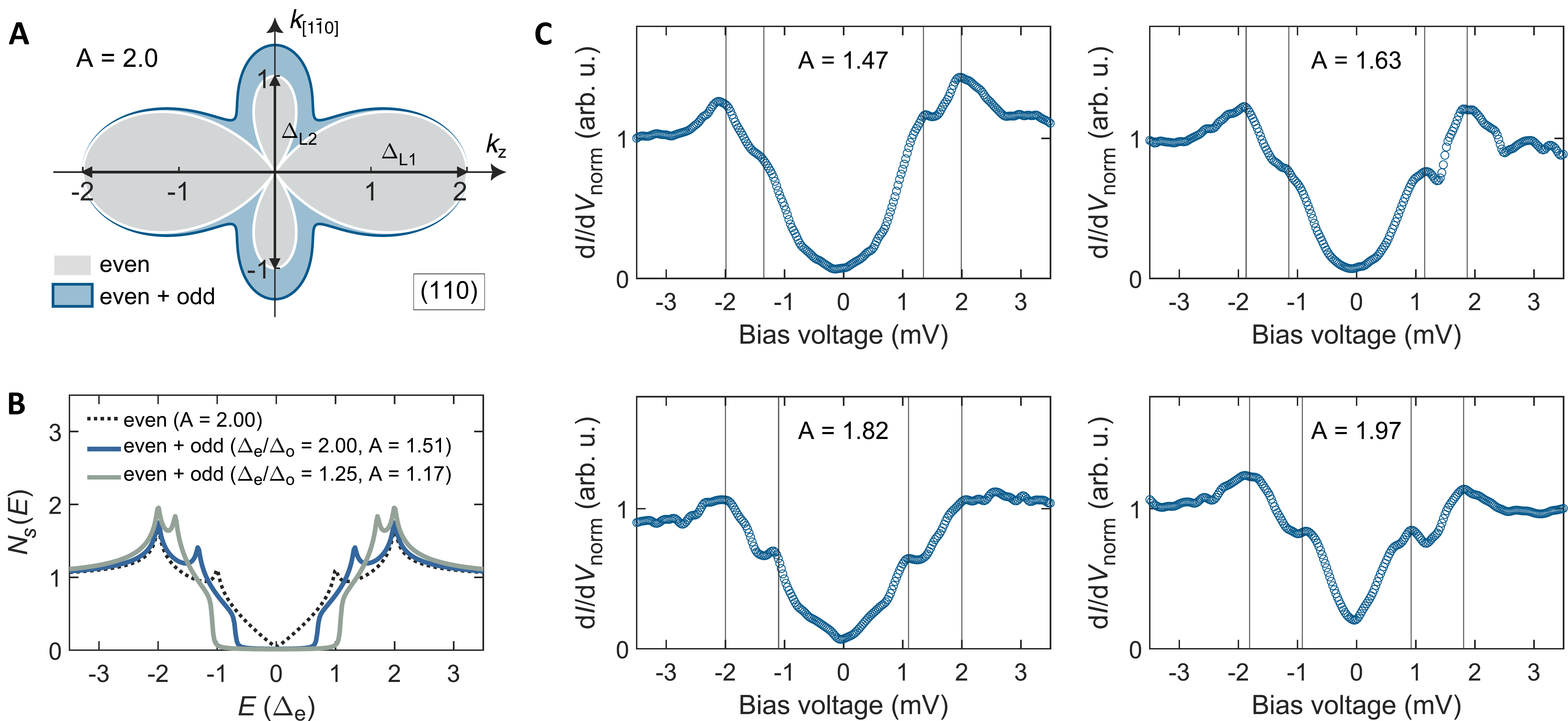}
    \caption{\textbf{Analysis of the enlarged order parameter acquired on the (110) surface orientation.} (\textbf{A}) Two-dimensional $k$-space cut through the even-parity $d_{z^2}$-orbital-type OP ($\Gamma_3$) along the (110) plane. Adding an odd-parity component ($\Gamma_3$) significantly alters the lobe structure. Particularly, the variable scale ratio $A=\Delta_{L1}/\Delta_{L2}$ of the orthogonally oriented lobes manifests as energy-dependent in-gap feature in the DoS. (\textbf{B}) Calculated DoS with variable contributions of the odd-parity component. (\textbf{C}) Comparison of experimental tunnel spectra acquired at four different positions on the (110) surface, revealing a correlation between the degree of its V-shape and the ratio $A$, as expected for the nodal pairing symmetry.}
    \label{Fig5}
\end{figure}

The physical picture drawn from this analysis is of two concurring effects: the nodal singlet component tends to produce V-shaped spectra, while the triplet component partially fills in the low-energy DoS, and the balance between the two components determines the spectral shape observed at any given location on the crystal surface. Also, for the enlarged spectra with double peaks, the ratio between $\Delta_e$ and $\Delta_o$ modulates the scale ratio $A=L_{1}/L_{2}$ of the orthogonally oriented lobes, and thereby the 
position of the inner peaks while the outer peaks remain unchanged (Figs.~\ref{Fig5}A-C). 

Fitting the measured d$I$/d$V$ spectra also allows us to determine the local singlet-to-triplet ratio (see Supplementary Text for details). Across the full dataset,  the amplitude of the triplet component ranges from 0\% $-$ corresponding to a superconducting OP containing only 
spin-singlet component $-$ to 40–50\% relative to the singlet part (as shown in Fig. \ref{Fig5} and Supplementary Fig. S8). The spatial variation of this ratio most likely reflects local inhomogeneities in the surface structure and composition, which modulate the local ASOC strength and hence the degree of parity mixing.

This mixed-parity superconducting OP also accounts for the anisotropic $B$-response observed on the (2$\overline{1}$2) surface (Fig.~\ref{Fig4}C) and allows us to infer the orientation of $\mathbf{d}(\mathbf{k})$. A $B$ applied parallel to $\mathbf{d}(\mathbf{k})$ should in fact preferentially suppress the triplet component, enhancing the relative contribution of the nodal singlet state and thereby producing a more pronounced V-shaped DoS. This behavior is observed experimentally for an in-plane $B$, $B_\parallel$, implying that the $\mathbf{d}$ vector has predominantly an in-plane orientation for the (2$\overline{1}$2)-oriented crystals (see Supplementary Text for details).

The complete orientation dependent spectroscopic dataset therefore identifies a mixed-parity superconducting state consisting of a nodal spin-singlet order parameter and a substantial spin-triplet component. The excellent agreement between experiment and theory demonstrates that ASOC alone is sufficient to generate strong parity mixing in the weakly correlated NCS Nb-Re, thereby identifying the origin of the mixed-parity character of its superconducting state.

\section*{Discussion}\label{discussion}

\noindent Our demonstration that a weakly correlated metal hosts a significant spin-triplet component, reaching up to half of the spin-singlet amplitude, shows that ASOC {\em per se} is sufficient to foster strong parity mixing. This result addresses a long-standing question in the physics of NCSs, where the most compelling evidence for mixed-parity pairing has historically emerged from heavy-fermion compounds 
such as CePt$_{3}$Si \cite{Bauer2004, Yogi2004, Yogi2006}, CeRhSi${_3}$ \cite{Kimura2005} and CeIrSi${_3}$ \cite{Mukuda2008, Tada2008}, in which strong correlations, flat electronic bands, and competing electronic orders obscure the specific role of ASOC. 

Half-Heusler compounds have likewise emerged as promising candidates for mixed-parity superconductivity through their $j=3/2$ fermions \cite{Ishihara2021}. In these materials, the interaction between strong spin-orbit coupling, multiband electronic structure and non-trivial topological bands complicates the identification of the microscopic origin of the superconducting state \cite{Brydon2016, Timm2017}.


By contrast, Nb-Re is a comparatively simple 
NCS \cite{Karki2011}, in which the consequences of ASOC can be identified more directly without the complexities associated with heavy-fermion correlations or multiband topology. The weakly correlated character of Nb-Re is also reflected in its superconducting length scales. For the Nb-Re single crystals studied here, using the Fermi velocity $v_\text{F} \sim 1.5 \times 10^5$ m/s obtained from DFT calculations (Fig.~\ref{Fig1}B) and the measured $\Delta \sim 1.1$\,meV, we estimate an upper bound for the superconducting coherence length $\xi_0 = \hbar v_\text{F}/\pi \Delta$ of $\sim 25\text{--}30\,\mathrm{nm}$, which places Nb-Re samples in the 
regime where Cooper pairs extend over several lattice constants. Despite its weakly correlated nature our analysis shows that the spin-triplet component in Nb-Re reaches nearly half of the spin-singlet amplitude, comparable to values reported for heavy-fermion superconductors
\cite{Kneidinger2015, Smidman2017}. The substantial size of the spin-triplet component is further supported by the large ratio $E_r = E_\text{ASOC}/k_\text{B} T_c$ $\approx 130$ ($E_\text{ASOC} \sim 100$\,meV based on the band splitting calculated with DFT and $T_c=8.8$\,K; ref. \cite{Jiao2014}), which falls within the range associated with substantial parity mixing in NCSs \cite{Jiao2014}. The estimated $E_r$ hence confirms that Nb-Re sits well within the strong parity-mixing regime, despite its weak electronic correlations.

Unlike heavy-fermion NCSs such as CeRhSi${_3}$ \cite{Kimura2005} and CeIrSi${_3}$ \cite{Mukuda2008, Tada2008}, where unconventional superconductivity emerges under high pressure, Nb-Re exhibits mixed-parity superconductivity at ambient pressure and comparatively higher $T_c$. 
The superconducting state of Nb-Re also differs from that of locally noncentrosymmetric systems such as the highly debated odd-parity candidate CeRh${_2}$As${_2}$, where unconventional multiphase superconductivity associated with staggered Rashba-type spin-orbit coupling from alternating locally noncentrosymmetric layers requires a high $B$ to be accessed \cite{Khim2021, Fischer2023}. By contrast, Nb-Re is a NCS with cubic symmetry, in which inversion-symmetry breaking gives rise to a more isotropic form of ASOC than in polar tetragonal systems such as Ce$M$Si$_3$ ($M$ = Rh, Ir), where ASOC is constrained to a single polar axis. The spin splitting between bands induced by ASOC in Nb-Re is therefore expected to be less anisotropic than in Rashba-type systems.


Our results imply that mixed-parity superconductivity does not necessarily require high pressure or magnetic fields, but can emerge at ambient pressure and $B$ = 0, at sufficiently low temperature, in weakly correlated NCSs with sufficiently strong ASOC.


The mixed-parity superconducting OP identified here also provides a unified interpretation of the apparently contradictory experimental literature on Nb-Re. Previous reports of conventional BCS-like superconductivity \cite{Cirillo2016, Lue2011, Karki2011, Chen2013} are fully consistent with our observations on (001)- and (113)-oriented surfaces, which exhibit a 
BCS-like DoS. Conversely, signatures of unconventional superconductivity find a natural explanation within the mixed-parity superconducting state identified here. Our results hence show that these apparently conflicting observations do not arise from inconsistencies in the superconducting state of Nb-Re, but rather from the different sensitivities of the experimental probes used. Techniques that average over crystallographic orientations or probe non-local properties inevitably sample different projections of the $k$-dependent superconducting OP. By directly correlating tunneling spectra with crystallographic orientation, our measurements remove this ambiguity and provide a quantitative $k$-space reconstruction of the mixed-parity superconducting state.

More generally, they demonstrate that 
orientation dependent tunneling spectroscopy offers a general strategy for identifying mixed-parity superconductivity.

Several independent observations reported in this study support the conclusion that the spin-triplet component in Nb-Re is intrinsic. The symmetry-constrained analysis first identifies a nodal spin-singlet $\Gamma_3$ state as the only even-parity OP capable of reproducing the V-shaped low-energy DoS. 
Crucially, this state alone cannot account for the complete orientation dependent spectroscopic dataset, which requires the inclusion of a sizable spin-triplet component. The resulting mixed-parity superconducting OP also consistently explains the anisotropic $B$-field dependence observed on the (2$\overline{1}$2) surface. Together, these observations demonstrate that the substantial spin-triplet component is a robust feature of the superconducting state in Nb-Re rather than an artifact of the specific model used to analyze the data.

Our modeling intentionally adopts the simplest symmetry-constrained description capable of capturing the experimental observations. While it neglects the detailed multiband electronic structure predicted by DFT and the ASOC-induced helicity-band splitting, these approximations primarily affect the quantitative value of the singlet-to-triplet ratio rather than the qualitative conclusion that a substantial 
spin-triplet component is required to reproduce the measured tunneling spectra. Likewise, the precise phase relationship between the singlet and triplet components, and its possible connection to the time-reversal-symmetry breaking reported by $\mu$SR measurements \cite{Shang2018}, remain important questions for future investigation.


Unlike bulk- or orientation averaged techniques, such as Knight shift or penetration depth measurements, our approach directly gives access to the $k$-space structure of the superconducting OP, which in turn enables the identification of the $k$-space distribution of the spin-triplet component. Notably, the $k$-space structure of a superconducting OP has been experimentally resolved in only a handful of studies, typically in superconductors with simpler band structures than Nb-Re ~\cite{Sharma2020,Allan2012}.
We anticipate that this methodology will facilitate the identification of mixed-parity superconductivity in a broader range of noncentrosymmetric materials, 
such as NbSe$_2$, for which a mixed $s+f$ OP has been suggested but not conclusively established experimentally\cite{Cho2022, Galvis2018}. 

Beyond fundamental implications, the thin-film compatibility of Nb-Re and related Re-based compounds, together with the substantial intrinsic spin-triplet component established here, makes them attractive candidates for superconducting spintronics, where the combination of these superconductors with ferromagnets can enable the generation of dissipationless spin transport. More generally, our work motivates future studies aimed at resolving the evolution of the superconducting OP with $B$ and $T$, as well as at exploring directional spin transport in crystallographically-controlled superconductor/ferromagnet heterostructures based on Nb-Re thin films.

Taken together, our results establish weakly correlated NCSs as a realistic materials platform in which substantial intrinsic spin-triplet pairing can emerge through ASOC alone, opening new opportunities to explore and exploit spin-triplet superconductivity beyond the realm of strongly correlated electron systems.

\section*{Materials and Methods}

\subsection*{Scanning tunneling microscopy}

\noindent Low-temperature scanning tunneling spectroscopy (STS) was performed on normal metal/insulator/ superconductor (N-I-S) tunnel junctions to probe the local density of states (DoS) of the superconductor. The measurement setup consists of a custom-built scanning tunneling microscope (STM), which is installed into a $^3$He cryostat operating at a base temperature of 280\,mK. A PtIr tip is attached to a piezo tube via an inertial slider design, allowing coarse and fine approach towards the sample by slip-stick motion \cite{Debuschewitz2007, Wolz2011}. All operations are controlled by a commercial SPM1000 STM controller (from company RHK), combined with a RHK IVP-300 current amplifier. At room temperature, an additional homemade voltage divider (ratio 1:100) is integrated into the circuit to improve the voltage resolution. To acquire differential conductance (d$I$/d$V$) spectra, the system utilizes a lock-in technique, with the STM's feedback control loop switched off during each voltage sweep. The corresponding alternating current modulation added to the direct current (DC) 
voltage biasing has an amplitude of $\sim 20\,\mu$V at a frequency of either 413 or 733\,Hz. The setpoint tunneling current was typically in the range of $0.4-1$\,nA, which, at bias voltages of a few mV, corresponds to a tunneling resistance of $1-10$\,M$\Omega$. The measurement parameters were regularly adjusted to maintain an optimal signal-to-noise ratio under varying experimental conditions. At base temperature, the setup is capable of performing STS with about 20\,$\mu$eV resolution.

\subsection*{Thin-film growth}

\noindent The polycrystalline \ce{Nb18Re82} thin films of 20\,nm thickness were grown by magnetron sputtering using an ATC Orion 5 UHV multitarget system. High-quality Si/\ce{SiO2} wafers, serving as substrates, were cut into $\sim 4 \times 4$\,mm$^2$ pieces to fit the dimensions of the STM sample holder. Prior to each deposition, the substrates underwent an ultrasonic cleaning in acetone and isopropanol and were subsequently dried with nitrogen gas. As an additional preparatory step, a brief pre-sputtering of a Ta or Ti target was performed to reduce the oxygen content and stabilise the ultra-high vacuum (UHV) chamber at a base pressure of $\sim 2.7\times 10^{-6}$\,Pa. Each deposition sequence began with a brief radiofrequency 
Ar plasma cleaning of the substrate for 30\,s at 50\,W, immediately followed by the Nb-Re growth from a single stoichiometric target (99.95\% purity). The first batch of samples was fabricated by DC magnetron sputtering at a deposition rate of 0.25\,nm/s. The plasma was ignited at a power of 250\,W in an Ar atmosphere at a pressure of 0.4\,Pa. A second batch of samples was grown by RF sputtering at a power of 200\,W under the same Ar pressure of 0.4\,Pa, resulting in a slightly 
smaller deposition rate of 0.09\,nm/s.

\subsection*{Single-crystal preparation}

\noindent Single crystals of Nb$_{x}$Re$_{1-x}$ with $x = 0.18$ have been synthesized by the floating zone technique using a four mirror optical furnace with Xe arc lamps as described in detail in Ref. \cite{Singh2012}. Single crystal pieces suitable for our experiments, aligned along specific crystallographic orientations were cut-out and selected from the as-grown batch using X-ray Laue diffraction. STS measurements were carried out on two distinct crystals. The first crystal C1 exposes a (001) surface. With lateral dimensions of approximately 2.5\,mm by 2\,mm, it could be readily mounted onto the STM sample holder by visual alignment.

The second crystal C2 was used to investigate three additional surface orientations (see Supplementary Fig.~S1). This larger piece featured the (113) and (110) surfaces as its native orientations. The third (2$\overline{1}$2) orientation was later obtained by independent diamond polishing. To carry out this step, the crystal was mounted in a self-constructed two-segmented sample holder allowing a precise angular alignment. The orientation was determined from the crystallographic relationship of the two reference surfaces, from which the rotation angles to bring the (2$\overline{1}$2) surface parallel to the polishing plane were calculated. The crystal was secured using a small amount of wax and polished with a Buehler IsoMet Low Speed Precision Cutter. Initial diamond cutting removed the bulk material, followed by successive polishing steps using grinding wheels of decreasing roughness. Subsequently, the (2$\overline{1}$2) surface was manually polished with diamond sandpapers of 8\,$\mu$m, 3\,$\mu$m, and 1\,$\mu$m grit to further reduce roughness and residual contamination.

Prior to the installation into the STM, each crystal surface was cleaned by Ar ion milling using an ArBlade 5000 system (from company Hitachi) equipped with an ion gun for flat milling. A shallow incidence angle of $\alpha_\text{in}=30^\circ$ was chosen to primarily modify near-surface layers, removing contamination, oxide, and defects. The argon flow was set to 0.16\,cm$^3$/min, with typical milling duration of 2 to 5\,min. The acceleration voltage $V_\text{acc}$ ranged from 4 to 8\,kV, while the discharge voltage $V_\text{dis}$ was between 1.5 and 2\,kV, depending on the surface requirements. A combination of X-ray diffraction (XRD) and electron backscatter diffraction (EBSD) measurements was used to characterise the crystallographic orientation of the single-crystal surfaces. The EBSD characterisation of the \{110\} and \{2$\overline{1}$2\} planes, which are not shown in the main text, is summarised in Supplementary Fig. S1.

\subsection*{Dynes fitting of the superconducting gap}

\noindent The tunneling spectra measured by the STM are analyzed with the phenomenological Dynes model. In this well-established framework, the superconducting DoS is expressed as
\begin{equation}\label{eq:Dynes_formula}
N_s(E) = N_0(E_F)\;\text{Re} \left[ \frac{E - i \gamma}{\sqrt{(E - i \gamma)^2 - \Delta(\theta)^2}} \right]\;,
\end{equation}
where $\Delta$ represents the superconducting order parameter (OP) and $\gamma$ describes finite quasiparticle lifetimes due to pair-breaking effects or residual experimental broadening. In the simplest case, the OP is written as $\Delta(\theta)=\Delta_0 \cos\left( n \theta \right)$ with $n=$ 0, 1, or 2, corresponding to $s$, $p$, or $d$ wave 
symmetries, respectively. More complex combinations of different angular-momentum pairing symmetries (mixed states) can also occur. The tunneling differential conductance d$I$/d$V(V)$ of a normal metal/superconductor tunnel  
junction at finite temperature $T$ is then obtained by convolving $N_s(E)$ with the derivative of the Fermi function $f$ and numerical integration via

\begin{equation}\label{eq:Dynes_dIdV}
\frac{\text{d}I}{\text{d}V} \propto \int_{-\infty}^{\infty} \text{d}E \int_0^{2\pi} \text{d}\theta\; N_s(E)  \; \left(-\frac{\partial f(E - eV, T)}{\partial E}\right) \vspace{0.5cm}
\end{equation}

The second integration over $\theta$ is necessary to account for the two-dimensional $k$-space averaging over the anisotropic Fermi surface in the case $n \geq 1$. Methodologically, Eq.~\eqref{eq:Dynes_dIdV} serves as a fit function to the experimental tunneling spectra.

For materials exhibiting multiple superconducting gaps with variably shaped OPs (multiband superconductivity), the total DoS can be represented as a weighted sum of individual contributions from each band, as given by

\begin{equation}
N_s(E) = \sum_{i} w_i \; N_{s,i}(E, \theta, \Delta_i, \gamma_i)\;.
\label{eq:two_gap}
\end{equation}

Here, $w_i$ with $\sum_i w_i = 1$ denotes the spectral weight of band $i$, associated with its band-specific tunnelling probability. The variables $\Delta_i$ and $\gamma_i$ are the corresponding OPs 
and Dynes broadening parameters. Note that the formalism as introduced above considers non-interacting bands with an intrinsic OP $\Delta_i$, originating from electron-phonon coupling in band $i$. We do not consider interband scattering as described by extended versions of the Dynes formula (McMillan model) \cite{Tao2022}.

\subsection*{Band-structure calculations}

\noindent For the computations, we employed density functional theory (DFT) in the generalized gradient approximation (GGA) with the Perdew-Burke-Ernzerhof~\cite{perdew_generalized_1996,perdew_generalized_1997} functional as implemented in the \texttt{VASP} 6.3.0 package~\cite{kresse_ab_1993,kresse_ab_1994,kresse_efficiency_1996,kresse_efficient_1996} within the projector augmented wave method (PAW)~\cite{blochl_projector_1994,kresse_ultrasoft_1999};
\verb|Nb_sv| and \verb|Re| pseudopotentials were used.
The energy cutoff was set to 550~eV, and a regular $\Gamma$-centered mesh with $R_k$ length of 50 was used.

The calculations were done for \ce{Nb17Re83} using an ordered structure and 20 representative structures aimed at modeling a substitutional disorder, each obtained by a random permutation of atoms of the ordered structure within the conventional unit cell.
This procedure was employed because it is not computationally feasible to explore the full space of possible representative structures.
Atomic positions within a fixed unit cell were optimized with the 1~meV/\r{A} convergence threshold for the absolute value of forces.

Two sets of calculations were done: with and without spin-orbit coupling taken into account.
The results in the main text are for the case with spin-orbit coupling included.
In both cases, the ordered structure was the one with the lowest energy and was thus selected as a representative for further analysis. The ordered structure is described by the space group \#217 with Nb occupying the (2a) and (2c) Wyckoff positions and Re occupying two (24g) Wyckoff positions.
The lattice parameters were taken from Ref.~\cite{Karki2011}.

The \texttt{ifermi} package~\cite{Ganose2021} was used to plot the Fermi surface and extract the Fermi velocities. 
The $k$-points labeling and coordinates were taken from the Bilbao Crystallographic Server~\cite{aroyo_brillouin_2014}.
The \texttt{VASPKIT} package~\cite{wang_vaspkit_2021} was used to aid with the post-processing of the \texttt{VASP} calculations. The dataset needed to reproduce and verify the computational results presented in this manuscript is publicly available at the TU Wien Research Data repository~\cite{repository}.


\newpage

\newpage

\newpage

\newpage


\clearpage 

%
\bibliography{science_template} 

@article{Debuschewitz2007,
  title={A compact and versatile scanning tunnelling microscope with high energy resolution for use in a $^3${He} cryostat},
  author={Debuschewitz, Christian and M{\"u}nstermann, Frank and Kunej, Vojko and Scheer, Elke},
  journal={J. Low Temp. Phys.},
  volume={147},
  pages={525--535},
  year={2007},
  publisher={Springer},
}

@article{Galvis2018,
  title={Tilted vortex cores and superconducting gap anisotropy in 2H-NbSe2},
  author={Galvis, JA and Herrera, E and Berthod, C and Vieira, S and Guillam{\'o}n, I and Suderow, H},
  journal={Commun. Phys.},
  volume={1},
   pages={30},
  year={2018}
 }

@article{Fischer2023,
   author = "Fischer, Mark H. and Sigrist, Manfred and Agterberg, Daniel F. and Yanase, Youichi",
   title = "Superconductivity and Local Inversion-Symmetry Breaking", 
   journal= "Annual Review of Condensed Matter Physics",
   year = "2023",
   volume = "14",
   pages = "153-172",
   publisher = "Annual Reviews"
}

@article{Tada2008,
author = {Tada ,Yasuhiro and Kawakami ,Norio and Fujimoto ,Satoshi},
title = {Microscopic Mechanism and Pairing Symmetry of Superconductivity in the Noncentrosymmetric Heavy Fermion Systems {CeRhSi$_3$} and {CeIrSi$_3$}},
journal = {Journal of the Physical Society of Japan},
volume = {77},
number = {5},
pages = {054707},
year = {2008},
eprint = {https://%doi.org/10.1143/JPSJ.77.054707}
}

@article{Timm2017,
  title = {Inflated nodes and surface states in superconducting half-Heusler compounds},
  author = {Timm, C. and Schnyder, A. P. and Agterberg, D. F. and Brydon, P. M. R.},
  journal = {Phys. Rev. B},
  volume = {96},
  issue = {9},
  pages = {094526},
  numpages = {16},
  year = {2017},
  publisher = {American Physical Society}
}

@article{Brydon2016,
  title = {Pairing of $j=3/2$ Fermions in Half-Heusler Superconductors},
  author = {Brydon, P. M. R. and Wang, Limin and Weinert, M. and Agterberg, D. F.},
  journal = {Phys. Rev. Lett.},
  volume = {116},
  issue = {17},
  pages = {177001},
  numpages = {5},
  year = {2016},
  publisher = {American Physical Society}
}

@article{Ishihara2021,
  title = {Tuning the Parity Mixing of Singlet-Septet Pairing in a Half-Heusler Superconductor},
  author = {Ishihara, K. and Takenaka, T. and Miao, Y. and Mizukami, Y. and Hashimoto, K. and Yamashita, M. and Konczykowski, M. and Masuki, R. and Hirayama, M. and Nomoto, T. and Arita, R. and Pavlosiuk, O. and Wi\ifmmode \acute{s}\else \'{s}\fi{}niewski, P. and Kaczorowski, D. and Shibauchi, T.},
  journal = {Phys. Rev. X},
  volume = {11},
  issue = {4},
  pages = {041048},
  numpages = {12},
  year = {2021},
   publisher = {American Physical Society}
}

@article{Mukuda2008,
  title = {Enhancement of Superconducting Transition Temperature due to the Strong Antiferromagnetic Spin Fluctuations in the Noncentrosymmetric Heavy-Fermion Superconductor ${\mathrm{CeIrSi}}_{3}$: A $^{29}\mathrm{Si}$ NMR Study under Pressure},
  author = {Mukuda, H. and Fujii, T. and Ohara, T. and Harada, A. and Yashima, M. and Kitaoka, Y. and Okuda, Y. and Settai, R. and Onuki, Y.},
  journal = {Phys. Rev. Lett.},
  volume = {100},
  issue = {10},
  pages = {107003},
  numpages = {4},
  year = {2008},
  publisher = {American Physical Society}
}

@article{Kimura2005,
  title = {Pressure-Induced Superconductivity in Noncentrosymmetric Heavy-Fermion ${\mathrm{CeRhSi}}_{3}$},
  author = {Kimura, N. and Ito, K. and Saitoh, K. and Umeda, Y. and Aoki, H. and Terashima, T.},
  journal = {Phys. Rev. Lett.},
  volume = {95},
  issue = {24},
  pages = {247004},
  numpages = {4},
  year = {2005},
  publisher = {American Physical Society}
}

@article{Khim2021,
author = {S. Khim  and J. F. Landaeta  and J. Banda  and N. Bannor  and M. Brando  and P. M. R. Brydon  and D. Hafner  and R. Küchler  and R. Cardoso-Gil  and U. Stockert  and A. P. Mackenzie  and D. F. Agterberg  and C. Geibel  and E. Hassinger },
title = {Field-induced transition within the superconducting state of {CeRh$_2$As$_2$}},
journal = {Science},
volume = {373},
number = {6558},
pages = {1012-1016},
year = {2021}
}

@article{Jiao2014,
  title = {Anisotropic superconductivity in noncentrosymmetric BiPd},
  author = {Jiao, L. and Zhang, J. L. and Chen, Y. and Weng, Z. F. and Shao, Y. M. and Feng, J. Y. and Lu, X. and Joshi, B. and Thamizhavel, A. and Ramakrishnan, S. and Yuan, H. Q.},
  journal = {Phys. Rev. B},
  volume = {89},
  issue = {6},
  pages = {060507(R)},
  numpages = {5},
  year = {2014},
   publisher = {American Physical Society}
}

@article{Yuan2006,
  title = {$S$-Wave Spin-Triplet Order in Superconductors without Inversion Symmetry: ${\mathrm{Li}}_{2}{\mathrm{Pd}}_{3}\mathrm{B}$ and ${\mathrm{Li}}_{2}{\mathrm{Pt}}_{3}\mathrm{B}$},
  author = {Yuan, H. Q. and Agterberg, D. F. and Hayashi, N. and Badica, P. and Vandervelde, D. and Togano, K. and Sigrist, M. and Salamon, M. B.},
  journal = {Phys. Rev. Lett.},
  volume = {97},
  issue = {1},
  pages = {017006},
  numpages = {4},
  year = {2006},
   publisher = {American Physical Society},
}

@article{Cho2022,
  title = {Nodal and Nematic Superconducting Phases in ${\mathrm{NbSe}}_{2}$ Monolayers from Competing Superconducting Channels},
  author = {Cho, Chang-woo and Lyu, Jian and An, Liheng and Han, Tianyi and Lo, Kwan To and Ng, Cheuk Yin and Hu, Jiaqi and Gao, Yuxiang and Li, Gaomin and Huang, Mingyuan and Wang, Ning and Schmalian, J\"org and Lortz, Rolf},
  journal = {Phys. Rev. Lett.},
  volume = {129},
  issue = {8},
  pages = {087002},
  numpages = {6},
  year = {2022},
   publisher = {American Physical Society},
}

@article{Yazdani1997,
author = {Ali Yazdani  and B. A. Jones  and C. P. Lutz  and M. F. Crommie  and D. M. Eigler },
title = {Probing the Local Effects of Magnetic Impurities on Superconductivity},
journal = {Science},
volume = {275},
pages = {1767-1770},
year = {1997},
}

@article{aroyo_brillouin_2014,
author = "Aroyo, Mois I. and Orobengoa, Danel and de la Flor, Gemma and Tasci, Emre S. and Perez-Mato, J. Manuel and Wondratschek, Hans",
title = "{Brillouin-zone database on the {\it Bilbao Crystallographic Server}}",
journal = "Acta Crystal. A",
year = "2014",
volume = "70",
pages = "126--137"
}

@article{Kashiwaya1995,
  title = {Origin of zero-bias conductance peaks in high-${\mathit{T}}_{\mathit{c}}$ superconductors},
  author = {Kashiwaya, Satoshi and Tanaka, Yukio and Koyanagi, Masao and Takashima, Hiroshi and Kajimura, Koji},
  journal = {Phys. Rev. B},
  volume = {51},
  issue = {2},
  pages = {1350--1353},
  numpages = {0},
  year = {1995},
    publisher = {American Physical Society},
}

@article{Hu1994,
  title = {Midgap surface states as a novel signature for {$d_{xa}^{2}$-$x_{b}^{2}$}-wave superconductivity},
  author = {Hu, Chia-Ren},
  journal = {Phys. Rev. Lett.},
  volume = {72},
  issue = {10},
  pages = {1526--1529},
  numpages = {0},
  year = {1994},
   publisher = {American Physical Society},
}

@article{Wolz2011,
  title = {Evidence for attractive pair interaction in diffusive gold films deduced from studies of the superconducting proximity effect with aluminum},
  author = {Wolz, M. and Debuschewitz, C. and Belzig, W. and Scheer, E.},
  journal = {Phys. Rev. B},
  volume = {84},
  issue = {10},
  pages = {104516},
  numpages = {10},
  year = {2011},
   publisher = {American Physical Society},
}

@article{Tao2022,
  title = {Multiband superconductivity in strongly hybridized 1{T'-WTe}$_{2}$/ {NbSe}$_{2}$ heterostructures},
  author = {Tao, Wei and Tong, Zheng Jue and Das, Anirban and Ho, Duc-Quan and Sato, Yudai and Haze, Masahiro and Jia, Junxiang and Que, Yande and Bussolotti, Fabio and Goh, K. E. Johnson and Wang, BaoKai and Lin, Hsin and Bansil, Arun and Mukherjee, Shantanu and Hasegawa, Yukio and Weber, Bent},
  journal = {Phys. Rev. B},
  volume = {105},
  issue = {9},
  pages = {094512},
  numpages = {14},
  year = {2022},
   publisher = {American Physical Society},
}

@article{Singh2012,
title = {Crystal growth of the non-centrosymmetric superconductor {Nb}$_{0.18}${Re}$_{0.82}$},
journal = {J. Crystal Growth},
volume = {361},
pages = {129-131},
year = {2012},
issn = {0022-0248},
author = {R.P. Singh and M. Smidman and M.R. Lees and D.McK Paul and G. Balakrishnan}
}

@article{Cirillo2025,
  title = {Unveiling Intrinsic Triplet Superconductivity in Noncentrosymmetric {NbRe} through Inverse Spin-Valve Effects},
  author = {Colangelo, F. and Modestino, M. and Avitabile, F. and Galluzzi, A. and Kakhaki, Z. Makhdoumi and Kumar, Abhishek and Linder, J. and Polichetti, M. and Attanasio, C. and Cirillo, C.},
  journal = {Phys. Rev. Lett.},
  volume = {135},
  issue = {22},
  pages = {226002},
  numpages = {6},
  year = {2025},
   publisher = {American Physical Society},
}

@article{Cirillo2015,
  title = {Evidence of double-gap superconductivity in noncentrosymmetric ${\mathrm{Nb}}_{0.18}{\mathrm{Re}}_{0.82}$ single crystals},
  author = {Cirillo, C. and Fittipaldi, R. and Smidman, M. and Carapella, G. and Attanasio, C. and Vecchione, A. and Singh, R. P. and Lees, M. R. and Balakrishnan, G. and Cuoco, M.},
  journal = {Phys. Rev. B},
  volume = {91},
  issue = {13},
  pages = {134508},
  numpages = {7},
  year = {2015},
    publisher = {American Physical Society},
}

@article{Cirillo2016,
  title = {Superconducting properties of noncentrosymmetric ${\mathrm{{Nb}}}_{0.18}{\mathrm{{Re}}}_{0.82}$ thin films probed by transport and tunneling experiments},
  author = {Cirillo, C. and Carapella, G. and Salvato, M. and Arpaia, R. and Caputo, M. and Attanasio, C.},
  journal = {Phys. Rev. B},
  volume = {94},
  issue = {10},
  pages = {104512},
  numpages = {8},
  year = {2016},
   publisher = {American Physical Society},
}

@article{Shang2018,
  title = {Time-Reversal Symmetry Breaking in {Re}-Based Superconductors},
  author = {Shang, T. and Smidman, M. and Ghosh, S. K. and Baines, C. and Chang, L. J. and Gawryluk, D. J. and Barker, J. A. T. and Singh, R. P. and Paul, D. McK. and Balakrishnan, G. and Pomjakushina, E. and Shi, M. and Medarde, M. and Hillier, A. D. and Yuan, H. Q. and Quintanilla, J. and Mesot, J. and Shiroka, T.},
  journal = {Phys. Rev. Lett.},
  volume = {121},
  issue = {25},
  pages = {257002},
  numpages = {7},
  year = {2018},
    publisher = {American Physical Society},
}

@Article{Kakhaki2023,
author="Zahra Makhdoumi Kakhaki and Antonio Leo and Federico Chianese and Loredana Parlato and Giovanni Piero Pepe and Angela Nigro and Carla Cirillo and Carmine Attanasio",
title="Upper critical magnetic field in N{bRe} and {NbReN} micrometric strips",
journal="Beilstein J. Nanotechnol.",
year="2023",
volume="14",
pages="45-51",
issn="2190-4286",
}

@article{Lue2011,
  title = {Evidence for $s$-wave superconductivity in noncentrosymmetric {Re}$_{24}${Nb}$_{5}$ from $^{93}${Nb} {NMR} measurements},
  author = {Lue, C. S. and Su, T. H. and Liu, H. F. and Young, Ben-Li},
  journal = {Phys. Rev. B},
  volume = {84},
  issue = {5},
  pages = {052509},
  numpages = {4},
  year = {2011},
   publisher = {American Physical Society},
}

@article{Karki2011,
  title = {Physical properties of the noncentrosymmetric superconductor {Nb}${}_{0.18}${Re}${}_{0.82}$},
  author = {Karki, A. B. and Xiong, Y. M. and Haldolaarachchige, N. and Stadler, S. and Vekhter, I. and Adams, P. W. and Young, D. P. and Phelan, W. A. and Chan, Julia Y.},
  journal = {Phys. Rev. B},
  volume = {83},
  issue = {14},
  pages = {144525},
  numpages = {6},
  year = {2011},
    publisher = {American Physical Society},
}

@article{Chen2013,
  title = {BCS-like superconductivity in the noncentrosymmetric compounds {Nb}${}_{x}${Re}${}_{1\ensuremath{-}x}$ ($0.13\ensuremath{\le}x\ensuremath{\le}0.38$)},
  author = {Chen, J. and Jiao, L. and Zhang, J. L. and Chen, Y. and Yang, L. and Nicklas, M. and Steglich, F. and Yuan, H. Q.},
  journal = {Phys. Rev. B},
  volume = {88},
  issue = {14},
  pages = {144510},
  numpages = {6},
  year = {2013},
    publisher = {American Physical Society}
}

@article{Kushwaha2024,
  title = {Unconventional properties of the noncentrosymmetric superconductor ${\mathrm{{Re}}}_{8}\mathrm{{NbTa}}$},
  author = {Kushwaha, R. K. and Arushi and Sharma, S. and Srivastava, S. and Meena, P. K. and Pula, M. and Beare, J. and Gautreau, J. and Hillier, A. D. and Luke, G. M. and Singh, R. P.},
  journal = {Phys. Rev. B},
  volume = {109},
  issue = {17},
  pages = {174518},
  numpages = {9},
  year = {2024},
   publisher = {American Physical Society},
}

@article{Biswas2012,
  title = {Comparative study of the centrosymmetric and noncentrosymmetric superconducting phases of {Re}${}_{3}${W} using muon spin spectroscopy and heat capacity measurements},
  author = {Biswas, P. K. and Hillier, A. D. and Lees, M. R. and Paul, D. McK.},
  journal = {Phys. Rev. B},
  volume = {85},
  issue = {13},
  pages = {134505},
  numpages = {7},
  year = {2012},
   publisher = {American Physical Society},
}

@article{Singh2017,
  title = {Time-reversal symmetry breaking in the noncentrosymmetric superconductor ${\mathrm{{Re}}}_{6}\mathrm{{Hf}}$: Further evidence for unconventional behavior in the $\ensuremath{\alpha}$-{Mn} family of materials},
  author = {Singh, D. and Barker, J. A. T. and Thamizhavel, A. and Paul, D. McK. and Hillier, A. D. and Singh, R. P.},
  journal = {Phys. Rev. B},
  volume = {96},
  issue = {18},
  pages = {180501(R)},
  numpages = {5},
  year = {2017},
   publisher = {American Physical Society},
}

@article{Singh2018,
  title = {Time-reversal symmetry breaking in the noncentrosymmetric superconductor ${\mathrm{{Re}}}_{6}\mathrm{{Ti}}$},
  author = {Singh, D. and K. P., Sajilesh and Barker, J. A. T. and Paul, D. McK. and Hillier, A. D. and Singh, R. P.},
  journal = {Phys. Rev. B},
  volume = {97},
  issue = {10},
  pages = {100505(R)},
  numpages = {6},
  year = {2018},
   publisher = {American Physical Society},
}

@article{Koch2024,
  title={Gate-controlled supercurrent effect in dry-etched Dayem bridges of non-centrosymmetric niobium rhenium},
  author={Koch, Jennifer and Cirillo, Carla and Battisti, Sebastiano and Ruf, Leon and Kakhaki, Zahra Makhdoumi and Paghi, Alessandro and Gulian, Armen and Teknowijoyo, Serafim and De Simoni, Giorgio and Giazotto, Francesco and others},
  journal={Nano Research},
  volume={17},
   pages={6575--6581},
  year={2024},
  publisher={Springer},
}

@Article{DeChiara2026,
author="Francesco De Chiara and Zahra Makhdoumi Kakhaki and Francesco Avitabile and Francesco Colangelo and Abhishek Kumar and Carmine Attanasio and Carla Cirillo",
title="Fast vortex dynamics and relaxation times in {NbRe}-based heterostructures",
journal="Beilstein J. Nanotechnol.",
year="2026",
volume="17",
pages="292-302",
issn="2190-4286",
}

@article{Cirillo2024,
  title={Single photon detection up to 2 $\mu$m in pair of parallel microstrips based on {NbRe} ultrathin films},
  author={Cirillo, C and Ejrnaes, M and Ercolano, P and Bruscino, C and Cassinese, A and Salvoni, D and Attanasio, C and Pepe, GP and Parlato, L},
  journal={Sci. Rep.},
  volume={14},
   pages={20345},
  year={2024},
  publisher={Nature Publishing Group UK London},
}

@article{Bauer2004,
  title = {Heavy Fermion Superconductivity and Magnetic Order in Noncentrosymmetric {CePt}$_{3}${Si}},
  author = {Bauer, E. and Hilscher, G. and Michor, H. and Paul, Ch. and Scheidt, E. W. and Gribanov, A. and Seropegin, Yu. and No\"el, H. and Sigrist, M. and Rogl, P.},
  journal = {Phys. Rev. Lett.},
  volume = {92},
  issue = {2},
  pages = {027003},
  numpages = {4},
  year = {2004},
    publisher = {American Physical Society},
}

@article{Yogi2006,
author = {Yogi, Mamoru and Mukuda, Hidekazu and Kitaoka, Yoshio and Hashimoto, Shin and Yasuda, Takashi and Settai, Rikio and D. Matsuda, Tatsuma and Haga, Yoshinori and \ifmmode \bar{O}\else \={O}\fi{}nuki, Yoshichika and Rogl, Peter and Bauer, Ernst},
title = {Evidence for Novel Pairing State in Noncentrosymmetric Superconductor {CePt}$_3${Si}: $^{29}${Si-NMR Knight} Shift Study},
journal = {J. Phys. Soc. Jpn},
volume = {75},
pages = {013709},
year = {2006},
}

@article{Yogi2004,
  title = {Evidence for a Novel State of Superconductivity in Noncentrosymmetric {CePt}$_{3}${Si}: A $^{195}${Pt-NMR} Study},
  author = {Yogi, M. and Kitaoka, Y. and Hashimoto, S. and Yasuda, T. and Settai, R. and Matsuda, T. D. and Haga, Y. and \ifmmode \bar{O}\else \={O}\fi{}nuki, Y. and Rogl, P. and Bauer, E.},
  journal = {Phys. Rev. Lett.},
  volume = {93},
  issue = {2},
  pages = {027003},
  numpages = {4},
  year = {2004},
   publisher = {American Physical Society},
}

@article{Diesch2018,
  title={Creation of equal-spin triplet superconductivity at the {Al/EuS} interface},
  author={Diesch, Simon and Machon, Peter and Wolz, Michael and S{\"u}rgers, Christoph and Beckmann, Detlef and Belzig, Wolfgang and Scheer, Elke},
  journal={Nat Commun.},
  volume={9},
   pages={5248},
  year={2018},
}

@article{Linder2015,
	title = {Superconducting spintronics},
	volume = {11},
    pages={307-315},
	journal = {Nat. Phys.},
	author = {Linder, Jacob and Robinson, Jason},
		year = {2015},
}

@article{Bernardo2015,
  title={Signature of magnetic-dependent gapless odd frequency states at superconductor/ferromagnet interfaces},
  author={Di Bernardo, Angelo and Diesch, Simon and Gu, Yuanzhou and Linder, Jacob and Divitini, Giorgio and Ducati, Caterina and Scheer, Elke and Blamire, Mark G and Robinson, Jason WA},
  journal={Nature Commun.},
  volume={6},
   pages={8053},
  year={2015},
}

@article{Smidman2017,
year = {2017},
publisher = {IOP Publishing},
volume = {80},
pages = {036501},
author = {Smidman, M and Salamon, M B and Yuan, H Q and Agterberg, D F},
title = {Superconductivity and spin–orbit coupling in non-centrosymmetric materials: a review},
journal = {Rep. Prog. Phys.}
}

@article{Sigrist2007,
title = {Superconductivity in non-centrosymmetric materials},
journal = {J. Mag. Mag. Mater.},
volume = {310},
pages = {536-540},
year = {2007},
issn = {0304-8853},
author = {Manfred Sigrist and D.F. Agterberg and P.A. Frigeri and N. Hayashi and R.P. Kaur and A. Koga and I. Milat and K. Wakabayashi and Y. Yanase},
}

@article{Yip2014,
   author = "Yip, Sungkit",
   title = "Noncentrosymmetric Superconductors", 
   journal= "Ann. Rev. Cond. Matter Phys.",
   year = "2014",
   volume = "5",
     pages = "15-33",
}

@article{Kneidinger2015,
title = {Superconductivity in non-centrosymmetric materials},
journal = {Physica C: Supercon. Appl.},
volume = {514},
pages = {388-398},
year = {2015},
issn = {0921-4534},
author = {F. Kneidinger and E. Bauer and I. Zeiringer and P. Rogl and C. Blaas-Schenner and D. Reith and R. Podloucky}
}

@article{Bergeret2005,
  author = {Bergeret, F. S. and Volkov, A. F. and Efetov, K. B.},
  title = {Odd triplet superconductivity and related phenomena in superconductor-ferromagnet structures},
  journal = {Rev. Mod. Phys.},
  volume = {77},
  pages = {1321--1373},
  year = {2005},
}

@article{Qiu2021,
author = {Qiu, Dong and Gong, Chuanhui and Wang, SiShuang and Zhang, Miao and Yang, Chao and Wang, Xianfu and Xiong, Jie},
title = {Recent Advances in {2D} Superconductors},
journal = {Adv. Mater.},
volume = {33},
pages = {2006124},
year = {2021}
}

@article{Kezilebieke2020,
  title={Topological superconductivity in a van der {Waals} heterostructure},
  author={Kezilebieke, Shawulienu and Huda, Md Nurul and Va{\v{n}}o, Viliam and Aapro, Markus and Ganguli, Somesh C and Silveira, Orlando J and G{\l}odzik, Szczepan and Foster, Adam S and Ojanen, Teemu and Liljeroth, Peter},
  journal={Nature},
  volume={588},
   pages={424--428},
  year={2020},
  publisher={Nature Publishing Group UK London},
}

@article{Sato2017,
year = {2017},
publisher = {IOP Publishing},
volume = {80},
pages = {076501},
author = {Sato, Masatoshi and Ando, Yoichi},
title = {Topological superconductors: a review},
journal = {Rep. Prog. Phys.}
}

@article{kresse_ab_1993,
	title = {Ab initio molecular dynamics for liquid metals},
	volume = {47},
		journal = {Phys. Rev. B},
	author = {Kresse, G. and Hafner, J.},
		year = {1993},
	pages = {558--561},
}

@article{kresse_ab_1994,
	title = {Ab initio molecular-dynamics simulation of the liquid-metal–amorphous-semiconductor transition in germanium},
	volume = {49},
		journal = {Phys. Rev. B},
	author = {Kresse, G. and Hafner, J.},
		year = {1994},
	pages = {14251--14269}
}

@article{kresse_efficient_1996,
	title = {Efficient iterative schemes for ab initio total-energy calculations using a plane-wave basis set},
	volume = {54},
		journal = {Phys. Rev. B},
	author = {Kresse, G. and Furthm\"uller, J.},
		year = {1996},
	pages = {11169--11186}
}

@article{kresse_efficiency_1996,
	title = {Efficiency of ab-initio total energy calculations for metals and semiconductors using a plane-wave basis set},
	volume = {6},
	issn = {0927-0256},
		journal = {Comp. Mater. Sci.},
	author = {Kresse, G. and Furthm\"uller, J.},
	year = {1996},
	pages = {15--50}
}

@article{blochl_projector_1994,
	title = {Projector augmented-wave method},
	volume = {50},
		date = {2023-07-27},
	journal = {Phys. Rev. B},
	author = {Bl\"ochl, P. E.},
		year = {1994},
	pages = {17953--17979}
}

@article{kresse_ultrasoft_1999,
	title = {From ultrasoft pseudopotentials to the projector augmented-wave method},
	volume = {59},
		date = {2023-07-27},
	journal = {Phys. Rev. B},
	author = {Kresse, G. and Joubert, D.},
		year = {1999},
	pages = {1758--1775}
}

@article{perdew_generalized_1996,
	title = {Generalized {Gradient} {Approximation} {Made} {Simple}},
	volume = {77},
    date = {2023-07-27},
	journal = {Phys. Rev. Lett.},
	author = {Perdew, John P. and Burke, Kieron and Ernzerhof, Matthias},
	year = {1996},
	pages = {3865--3868},
}

@article{perdew_generalized_1997,
	title = {Generalized {Gradient} {Approximation} {Made} {Simple} [{Phys}. {Rev}. {Lett}. 77, 3865 (1996)]},
	volume = {78},
    date = {2023-07-27},
	journal = {Phys. Rev. Lett.},
	author = {Perdew, John P. and Burke, Kieron and Ernzerhof, Matthias},
	year = {1997},
	pages = {1396--1396},
}

@article{Ganose2021,
  year = {2021},
  publisher = {The Open Journal},
  volume = {6},
   pages = {3089},
  author = {Ganose, Alex M. and Searle, Amy and Jain, Anubhav and Griffin, Sinéad M.},
  title = {IFermi: A python library for {Fermi} surface generation and analysis},
  journal = {Journal of Open Source Software}
}

@misc{repository,
title={Numerical data for "Orientation-resolved tunnel spectroscopy of a triplet superconductor"},
publisher={TU Wien},
author={Strohmeier, M. and Smolyanyuk, A. and Held, K. and Smidman, M. and Balakrishnan, G. and Belzig, W. and Scheer, E. and Di Bernardo, A.},
year={2026},
}

@article{wang_vaspkit_2021,
title = {VASPKIT: A user-friendly interface facilitating high-throughput computing and analysis using VASP code},
journal = {{Comp. Phys. Commun.}},
volume = {267},
pages = {108033},
year = {2021},
issn = {0010-4655},
author = {Vei Wang and Nan Xu and Jin-Cheng Liu and Gang Tang and Wen-Tong Geng}
}

@book{Savitskii1966,
  author    = {Savitskii, E. M. and Baron, V. V. and Efimov, Yu. V. and Bychkova, M. I. and Myzenkova, L. F.},
  title     = {Superconducting Materials},
  publisher = {Plenum Press},
  address   = {New York--London},
  year      = {1966}
}

@article{Bao2013,
  author  = {Bao, G. and Inada, Y. and Eguchi, G. and Maeno, Y. and Ichioka, M. and Zheng, G.-q.},
  title   = {Influence on superconductivity in the parity mixing superconductor {L}i$_2${T}$_3${B}({T}:{P}t,{P}d) by non-magnetic impurity and defect doping},
  journal = {Physica C},
  volume  = {494},
  pages   = {95--98},
  year    = {2013}
}

@article{Xu2014,
  author  = {Xu, Guizhou and Wang, Wenhong and Zhang, Xiaoming and Du, Yin and Liu, Enke and Wang, Shouguo and Wu, Guangheng and Liu, Zhongyuan and Zhang, Xi Xiang},
  title   = {Weak Antilocalization Effect and Noncentrosymmetric Superconductivity in a Topologically Nontrivial Semimetal {L}u{P}d{B}i},
  journal = {Sci. Rep.},
  volume  = {4},
  pages   = {5709},
  year    = {2014}
}

@article{Kim2018,
  author  = {Kim, Hyunsoo and Wang, Kefeng and Nakajima, Yasuyuki and Hu, Rongwei and Ziemak, Steven and Syers, Paul and Wang, Limin and Hodovanets, Halyna and Denlinger, Jonathan D. and Brydon, Philip M. R. and Agterberg, Daniel F. and Tanatar, Makariy A. and Prozorov, Ruslan and Paglione, Johnpierre},
  title   = {Beyond triplet: Unconventional superconductivity in a spin-3/2 topological semimetal},
  journal = {Sci. Adv.},
  volume  = {4},
  pages   = {eaao4513},
  year    = {2018}
}

@article{Sharma2020,
  author  = {Sharma, R. and Fang, S. and Chen, X. and Jain, S. and Li, T. and Vig, S. and Kim, H. and Slager, R.-J. and Kirshenbaum, K. and Butch, N. P. and Ali, M. N. and Fu, L. and D{\'a}vila, M. E. and Ando, Y. and Berger, H. and Dresselhaus, M. S. and Wray, L. A.},
  title   = {Momentum-resolved superconducting energy gaps of {S}r$_2${R}u{O}$_4$ from quasiparticle interference imaging},
  journal = {Proc. Natl. Acad. Sci.},
  volume  = {117},
  pages   = {5222--5227},
  year    = {2020}
}

@article{Allan2012,
  author  = {Allan, M. P. and Rost, A. W. and Mackenzie, A. P. and Xie, Yang and Davis, J. C. and Kihou, K. and Lee, C. H. and Iyo, A. and Eisaki, H. and Chuang, T.-M.},
  title   = {Anisotropic Energy Gaps of Iron-Based Superconductivity from Intraband Quasiparticle Interference in {L}i{F}e{A}s},
  journal = {Science},
  volume  = {336},
  pages   = {563--567},
  year    = {2012}
}

@article{Datta2021,
  author  = {Datta, Soumya and Vasdev, Aastha and Ramachandran, Ranjani and Halder, Soumyadip and Motla, Kapil and Kataria, Anshu and Arushi and Chowdhury, Rajeswari Roy and Singh, Ravi Prakash and Sheet, Goutam},
  title   = {Spectroscopic evidence of mixed angular momentum symmetry in non-centrosymmetric {R}u$_7${B}$_3$},
  journal = {Sci. Rep.},
  volume  = {11},
  pages   = {21030},
  year    = {2021}
}
\bibliographystyle{sciencemag}

%
%
%
%
%
%


\section*{Acknowledgments}
We are deeply indebted to Manfred Sigrist for his valuable scientific input and insightful discussions. We also thank Maksym Serbyn, Antonio Vecchione, Lilia Boeri and Simone Di Cataldo for fruitful discussion.
\paragraph*{Funding:}
We acknowledge funding through the Spezialforschungsbereich (SFB) Q-M\&S
of the Austrian Science Fund (FWF, project DOI 10.55776/F86),
with the DFG-funded subproject 493158779.
This research was furthermore funded in part by the FWF through project 10.55776/I6142. 
The work at the University of Warwick was supported by EPSRC, UK, through grant EP/T005963/1.

\paragraph*{Author contributions:}
AdB, MSt and ES designed the experiment. MSt conducted the sample preparation and characterization, performed the STM experiments, and carried out the data analysis and modeling of the DoS under the supervision of AdB and ES. AS and KH performed the DFT calculations. The Nb-Re single crystals were grown and characterized for the measurements at Warwick by GB and MSc. All authors discussed the results. AdB and MSt wrote the manuscript with input from all authors.
\paragraph*{Competing interests:}
There are no competing interests to declare.
\paragraph*{Data, code and materials availability:}
Access to the experimental data is available upon reasonable request. The data needed to reproduce and verify the computational results is publicly available at the TU Wien Research Data repository~\cite{repository}.


\subsection*{Supplementary materials}
Supplementary Text\\
Figs. S1 to S11\\
References\\ 




\end{document}